\documentclass{article}
\usepackage{graphicx}

\date{~}

\def\fun{ {\cal F} } 

\begin{document}

\title{%
\vskip-18pt~\\
Double-field Inflation}

\author{Fred Adams\rlap{}{$^{1,3}$} and 
Katherine Freese\rlap{}{$^{2,3}$} 
  \\~\\
\small \it ${}^{1}$Harvard-Smithsonian Center for Astrophysics,
Cambridge, MA 02138,
  \\
\small \it ${}^{2}$Physics Dept., Massachusetts Institute of
Technology, Cambridge, MA 02139
  \\
\small \it ${}^{3}$Current Address:  Michigan Center for Theoretical Physics,
\\
\small \it Physics Dept., University of Michigan, Ann Arbor, MI 48109
  \\
  \vspace{-2\baselineskip} }

\maketitle
 
\begin{abstract} 
  \noindent

We present an inflationary universe model which utilizes two coupled
real scalar fields.  The inflation field $\phi$ experiences a first
order phase transition and its potential dominates the energy density
of the Universe during the inflationary epoch.  This field $\phi$ is
initially trapped in its metastable minimum and must tunnel through a
potential barrier to reach the true vacuum.  The second auxiliary
field $\psi$ couples to the inflaton field and serves as a catalyst to
provide an abrupt end to the inflationary epoch; i.e., the $\psi$
field produces a time-dependent nucleation rate for bubbles of true
$\phi$ vacuum.  In this model, we find that bubbles of true vacuum can
indeed percolate and we argue that thermalization of the interiors can
more easily take place.  The required degree of flatness (i.e., the
fine tuning) in the potential of the $\psi$ field is comparable to
that of other models which invoke slowly rolling fields. Pseudo
Nambu-Goldstone bosons may naturally provide the flat potential for
the rolling field.

\end{abstract}

\newpage

\section{Introduction}

In 1981, Guth \cite{guth} proposed the inflationary universe model
to solve several cosmological problems, notably the
horizon problem, the flatness problem, and the monopole
problem.  During the inflationary epoch,
the energy density of the Universe is dominated by a
(nearly constant) false vacuum energy term $\rho \simeq \rho_{vac}$
= {\it constant},
and the scale factor of the Universe expands exponentially:
\begin{equation}
H^2 = 8\pi G \rho /3 ,
\end{equation}
\begin{equation} 
R(t)= R(t_0) e^{\chi (t-t_0)} \, , 
\end{equation}
where $H$ = $\dot R /R$ is the Hubble parameter, $R$ is the
scale factor of the Universe, $R(t_0)$ is the scale factor at 
the beginning of inflation, and $\chi$ is defined by
\begin{equation}
\chi = \sqrt{8\pi G \rho_{vac}/3} \,  
\end{equation}
(notice that $\chi$ $\approx$ $H$ during the inflationary epoch).
During this period of exponential expansion, a small causally
connected region of the Universe inflates to a sufficiently large
region to explain the observed homogeneity and isotropy of the
Universe today, to `inflate away' the overdensity of monopoles to
regions outside our horizon, and to predict a flat Universe with
$\Omega = 1$.  A successful resolution to these cosmological problems
requires at least 70 $e$-folds of inflation, i.e., the scale factor must
increase by at least $10^{27}$ (for $\chi \sim {\rm const}$).  The
period of exponential expansion must be followed by a period of
thermalization, in which the vacuum energy density is converted to
radiation.

In the the original inflationary model \cite{guth} (now known as `old'
inflation), the Universe supercools to a temperature $T \ll T_c$
during a first order phase transition with critical temperature $T_c$.
The nucleation rate for bubbles of true vacuum must be slow enough
that the Universe remains in the metastable false vacuum long enough
for at least 70 $e$-folds of inflation. Unfortunately, the old
inflationary scenario has been shown to fail \cite{guthwein} because
the interiors of expanding spherical bubbles of true vacuum fail to
thermalize -- the `graceful exit' problem.  Hence this model does not
produce a Universe such as our own.  The problem of ending old
inflation will be discussed in greater detail in Sec. II, where we
discuss modifications which can lead to percolation and
thermalization.

Linde \cite{linde} and Albrecht and Steinhardt \cite{as} proposed the
`new' inflationary scenario, in which the effective potential (or free
energy) of the inflation field becomes very flat (the phase transition
may now be second order or only weakly first order).  As the field
$\psi$ `slowly rolls' down the potential, the evolution of the field
is described by
\begin{equation}
\label{eq:slowroll}
\ddot\psi + 3H\dot\psi + \Gamma\dot\psi
+ {dV \over d\psi} = 0\ . 
\end{equation}
In the slowly rolling regime of growth, the energy density of the
Universe is dominated by the vacuum contribution ($\rho \simeq
\rho_{vac}$ $\gg$ $\rho_{rad}$) and the Universe expands
exponentially.  When the field approaches the true vacuum, it
oscillates about the minimum, and the $\Gamma \dot\psi$ term gives
rise to particle and entropy production.  In this manner, a `graceful
exit' to inflation is achieved.  Many other proposed versions of
inflation (e.g., the `chaotic' inflation model of Linde
\cite{chaotic}) utilize a slowly rolling field.

All existing versions of inflation with rolling fields tend to
overproduce density fluctuations and are thus highly constrained by
isotropy measurements of the microwave background \cite{cmb}.  These
measurements indicate that the amplitude of the density perturbations
must be less than $\delta$ $\approx$ $10^{-5}$.  However, inflationary
models predict \cite{star,bar} density fluctuations with amplitudes given by
\begin{equation}
\label{eq:densfluc}
{\delta \rho \over \rho}\bigg|_{hor} \simeq 
{H^2 \over \dot\psi} ,
\end{equation}
where the right hand side is evaluated at the time when
the fluctuation crossed outside the horizon during inflation 
and where ${(\delta \rho /\rho)}|_{hor}$ is the amplitude
of a density perturbation when it crosses back inside
the horizon after inflation.  In order for sufficient 
inflation to take place and for the density perturbations
to be smaller than the observational limits, the potential 
of the rolling field must be very flat. This statement can
be quantified \cite{afg} by defining a fine tuning parameter 
$\lambda$ through 
\begin{equation}
\label{eq:finpar}
\lambda \equiv {\Delta V \over (\Delta \psi)^4 } , 
\end{equation}
where $\Delta V$ is the change in the total potential $V(\psi)$ which
affects the $\psi$ field (including any interaction terms) and $\Delta
\psi$ is the change in the field $\psi$ during the slowly rolling
portion of the inflationary epoch.  The parameter $\lambda$ is
constrained to be small [i.e., $\lambda$ $\le$ $\cal O$ ($10^{-8}$ --
$10^{-11}$) ] for a general class of inflationary scenarios which
contain a slowly rolling field \cite{afg}.

In Sec. II, we discuss the `graceful exit' problem of old inflation
and discuss a mechanism to circumvent this problem.  In old inflation,
a small nucleation rate (which is constant in time) allows for
sufficient inflation, but the phase transition can never be completed.
A large nucleation rate (also constant in time) would allow the phase
transition to complete, but the Universe would not inflate
sufficiently to solve the cosmological problems stated above.  The
basic feature of this present scenario is to have a {\it
time-dependent} nucleation rate for bubbles of true vacuum in a first
order transition.  This time dependence allows us to take advantage of
the best features of both slow and fast nucleation rates. In our
scenario, the nucleation rate is initially negligible and the Universe
can inflate; subsequently, at the same time at every point (in a large
enough region of space to encompass our Universe), the nucleation rate
suddenly becomes extremely fast and the phase transition completes.
In Sec. III, we discuss a particular model to obtain a time dependent
nucleation rate which can produce a fairly sudden end to the phase
transition.  In this model, the old inflationary field $\phi$ is
coupled to a slowly rolling field $\psi$ which evolves in a flat
potential (like in new inflation). The $\phi$ field dominates the
dynamics of the Universe and gives rise to an inflationary epoch.  The
purpose of the slowly rolling field is to give the $\phi$ field a
time-dependent nucleation rate. When the slowly rolling field $\psi$
approaches its vacuum expectation value (the minimum of the
potential), the interaction between the fields catalyses the old
inflationary field $\phi$ to rapidly nucleate bubbles of true vacuum
throughout space.  However, the rolling field produces density
fluctuations with the same amplitude as in new inflationary models,
and bubble interactions on much smaller scales probably cannot erase
these large-scale fluctuations; hence, this model suffers from a
fine-tuning problem similar to that of new inflation.  However, as
discussed in Ref. \cite{afg}, this fine tuning problem is a generic
feature of inflationary models with slowly rolling fields; a
resolution (natural inflation with pseudo Nambu-Goldstone bosons) is
suggested in Ref. \cite{nat}.  This present model -- double field
inflation -- thus remains a viable alternative scenario in which the
end of the inflationary epoch occurs through the nucleation of
bubbles.  Although other inflationary models which use more than one
scalar field have been proposed \cite{more}, this present model is
different in that it achieves successful inflation through a
time-dependent nucleation rate and hence a time-dependent nucleation
efficiency $\beta$ (see Eq.{\ref{eq:beta}).  Some of the advantages
and disadvantages of this model are discussed in Sec. IV. For example,
cosmic strings can be formed at the end of the (first-order) phase
transition by the inflaton field $\phi$.

\section{Basic Mechanism}

Here we review the reasons for the failure of old inflation and
present a possible mechanism to circumvent these problems. 
In order to use a simple but illustrative example, we 
consider a quantum field theory of a scalar field with 
a Lagrangian of the form 
\begin{equation}
\label{eq:lag}
{\cal L} = {1 \over 2}(\partial_{\mu}\phi)(\partial^{\mu} \phi) 
- V_1(\phi) , 
\end{equation} 
where $V_1 (\phi)$ is an asymmetric potential with
metastable minimum $\phi_-$ and absolute minimum $\phi_+$
(see Fig. 1).  The energy difference between the vacua
is $\epsilon$.  Bubbles of true vacuum ($\phi_+$) expand
into a false vacuum ($\phi_-$) background. 

In the zero temperature limit, the nucleation rate $\Gamma_N$ 
(per unit time per unit volume) for producing bubbles of true 
vacuum in the sea of false vacuum through quantum tunneling can 
be calculated \cite{callan,coleman} and has the form 
\begin{equation}
\label{eq:nucrate}
\Gamma_N (t) = A e^{-S_E} ,
\end{equation}
where $S_E$ is the Euclidean action \cite{coleman} corresponding to
Eq.(\ref{eq:lag}) and where $A$ is a determinantal factor \cite{callan}
which is generally of order $T_c^4$ (where $T_c$ is the energy scale
of the phase transition).  In old inflation, this nucleation rate is
taken to be approximately constant in time throughout the phase
transition.  Guth and Weinberg \cite{guthwein} have shown that the
probability of a point remaining in the false-vacuum phase during the
transition (which begins at $t_i$) is given by
\begin{equation}
\label{eq:prob}
p(t) = \exp \Biggl\{ - \int^t_{t_i}\ dt^\prime 
\Gamma_N (t^\prime) R^3(t^\prime) {4 \pi \over 3} \left[ 
\int^t_{t^\prime}\ {dt^{\prime\prime} \over R(t^{\prime\prime})}
\right]^3 \Biggr\} . 
\end{equation}
During the de Sitter phase of expansion, the exponent in Eq.(\ref{eq:prob})
is approximately $-(4/3) \pi \beta \chi(t-t_i)$, where 
the dimensionless quantity $\beta$ is defined by 
\begin{equation}
\label{eq:beta}
\beta \equiv { \Gamma_N \over \chi^4 } . 
\end{equation}
The value of this {\it nucleation efficiency} $\beta$ 
can be calculated from the potential and is crucial 
for determining the nature of the phase transition.  

\begin{figure}
\centerline{\includegraphics[width=5.5in]{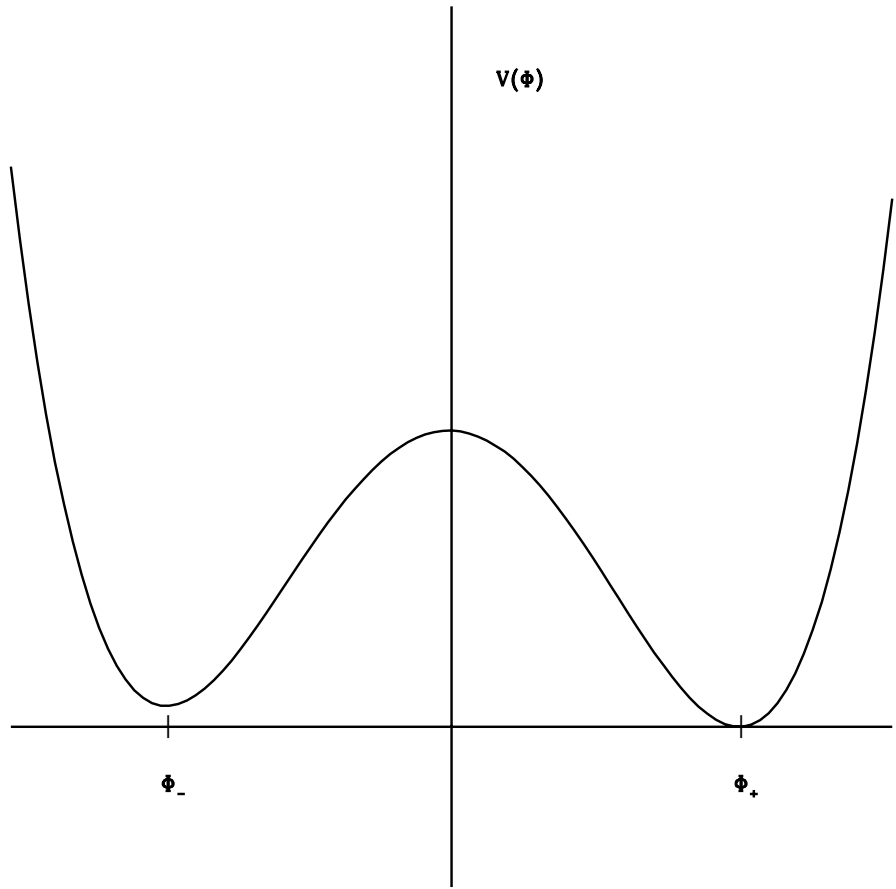}}
\caption{Potential energy density of inflaton field $\phi$ as a
function of field strength.  The energy difference $\epsilon$ between
the false vacuum (at $\phi_- = -a$) and the true vacuum (at $\phi_+ =a$)
provides the vacuum energy density for inflation.}
\end{figure}

In the limit that $\beta$ is small compared to unity (i.e., low
nucleation efficiency), the phase transition proceeds slowly and the
Universe can inflate through many $e$-foldings.  This limit corresponds
to the case of old inflation. However, when $\beta$ is sufficiently
small, the rate of filling the Universe with true vacuum cannot keep
up with the exponential expansion of the false vacuum and bubble
percolation never occurs, i.e., the phase transition is never
completed \cite{guthwein}.  In addition, thermalization of individual
bubbles or groups of bubbles never occurs.  Those bubbles which
nucleate early are quite large by the time later bubbles nucleate;
hence a wide distribution of bubble sizes is produced.  Groups of
bubbles are dominated by the single largest bubble in a cluster.  In
any single bubble, the latent heat of the phase transition
($\epsilon$) is entirely converted into the kinetic energy of the
bubble wall \cite{coleman} rather than thermalizing the interior of the bubble;
in addition, collisions with much smaller bubbles cannot thermalize
the interior.  As a result, the (nearly) homogeneous and isotropic
Universe we live in today can neither arise from a single large bubble
nor from clusters of bubbles.

In the opposite limit when $\beta$ is large compared to unity
(i.e., high nucleation efficiency), the phase transition 
proceeds very rapidly.  The timescale for bubbles to nucleate 
and percolate is small compared to the expansion timescale 
(which is determined by $\chi$) for the Universe.  In this 
limit, the phase transition is readily completed, but the 
Universe does not inflate sufficiently.  

A critical value $\beta_{CR}$ must exist \cite{guthwein}, such that
$\beta \ge \beta_{CR}$ implies percolation (the supercritical regime)
and $\beta \le \beta_{CR}$ implies no percolation (the subcritical
regime).  The critical value $\beta_{CR}$ lies in the range (see
Ref. \cite{guthwein})
\begin{equation}
0.24 \ge \beta_{CR} \ge 10^{-6} ,
\end{equation}
although alternate arguments \footnote{This range of values for
$\beta_{CR}$ values, including ``alternate arguments'', has been
discussed by E. Weinberg -- see Ref. \cite{weinberg}} have suggested
$\beta_{CR}$ $\approx$ 0.03.  As discussed above, $\beta$ must be
subcritical to allow for sufficient inflation {\it and} $\beta$ must
be supercritical to allow for percolation and hence to allow the phase
transition to complete.  Theories with constant $\beta$ (i.e., a
constant nucleation rate and a constant vacuum contribution $\chi$)
must clearly fail\footnote{This fact was noted by Guth in the original
paper -- see Ref. \cite{guth}}.

In this present model, we consider a nucleation rate 
(and hence $\beta$) which can vary with time.  
The nucleation rate is initially small 
(so that $\beta <$ $\beta_{CR}$). 
The Universe remains in the false vacuum and inflates for 
a long time.  As the Universe evolves, the nucleation rate
grows, and eventually $\beta$ becomes supercritical.  The 
bubbles of true vacuum can then percolate and the phase 
transition can be completed.  As long as the timescale
for $\beta$ to evolve from a subcritical value to a 
supercritical value is long enough to allow for sufficient
expansion of the Universe, a successful inflationary 
epoch will arise.  

For definiteness, we take the potential of 
the inflation field to be 
\begin{equation}
\label{eq:phipot}
V_1(\phi) = {1 \over 8} \lambda(\phi^2 - a^2)^2 - 
{\epsilon \over 2a} (\phi - a) . 
\end{equation}
To leading order, the metastable minimum is given by
$\phi_- = -\ a$ and the absolute minimum by $\phi_+ = +\ a$.
In addition, we will take an interaction term of the form
\begin{equation}
V_{int}(\phi) = - {1 \over 2 a} Y(\psi) 
a^4 (\phi - a) \ , 
\end{equation}
where $Y$ is a dimensionless function which evolves in time and is 
independent of the $\phi$ field.  In this case, the 
effective energy difference between the vacua 
(see $\epsilon$ in Eq.(\ref{eq:phipot}) is given by 
\begin{equation}
\label{eq:eff}
\epsilon_{eff} =  \epsilon + Y(\psi) a^4 .  
\end{equation}
Bubbles will nucleate at a rate given by Eq. (\ref{eq:nucrate}).  For the 
potential of Eq. (\ref{eq:phipot}) and in the limit that the nondegeneracy of 
the vacua is small (i.e., $\epsilon$ small), the Euclidean action 
can be obtained analytically \cite{coleman} and is given by 
\begin{equation}
\label{eq:eucaction}
S_E = {\pi^2 \over 6} {\lambda^2 a^{12} \over
{\epsilon_{eff}^3}} . 
\end{equation}
The limit of small $\epsilon$ is sometimes denoted as 
``the thin wall limit'' because the validity of the 
the analytic expression (\ref{eq:eucaction}) is limited to cases 
in which the wall thickness of the nucleated bubble
is small compared to the bubble radius (see Ref. \cite{coleman}). 
If the function $Y$ changes from a very small initial value 
(which leads to a small nucleation rate) to a large value 
at some later time $t_f$ throughout space, a large nucleation 
rate will result and the phase transition can come to completion 
near the time $t_f$.  If the end of this phase transition is 
sufficiently abrupt, bubbles of nearly equal size will nucleate 
simultaneously everywhere in space. Thus both percolation and 
thermalization can be more easily achieved. Any cluster of bubbles 
consists of equal-sized bubbles which can more easily thermalize 
one another than the wide variety of bubble sizes arising in old 
inflation. 

\section{Double Field Inflation}

In this section, we implement the ideas discussed in the previous 
section by presenting a particular model in which the interaction 
term is given by the interaction of the inflation field $\phi$ with 
a second scalar field; the potential of this second field $\psi$ 
is very flat and gives rise to slowly rolling behavior, just as 
in new inflation. The old inflation field $\phi$, which is initially 
trapped in its metastable minimum and must tunnel through a potential 
barrier, dominates the dynamics and causes the Universe to inflate.  
The rolling field merely serves as a catalyst for an abrupt end to 
the inflationary epoch, i.e., the $\psi$ field produces the desired 
time-dependent nucleation rate for bubbles of true $\phi$ vacuum. 
In this model, we find that bubbles of true vacuum can indeed percolate 
and we argue that thermalization of the interiors can more easily 
take place. 

\subsection{The Model}

The total Lagrangian (for both fields) has the form 
\begin{equation}
\label{eq:totlag}
{\cal L}={1 \over 2} (\partial_{\mu}\phi)^2 + {1 \over 2}
(\partial_{\mu}\psi)^2 - V_{tot}(\phi,\psi) , 
\end{equation}
where the total potential can be written 
\begin{equation}
\label{eq:totpot}
V_{tot}(\phi,\psi)=V_1(\phi) + V_2(\psi) + 
V_{int}(\phi,\psi)\ .
\end{equation}
For the sake of definiteness, we take $V_1(\phi)$ to be the 
potential of Eq. (\ref{eq:phipot}).  The potential $V_2(\psi)$ can be 
any flat potential which leads to slow rolling behavior of 
the $\psi$ field [see Eq.(\ref{eq:slowroll})].  For 
convenience, we take the interaction term to be of the form 
\begin{equation}
\label{eq:intpot}
V_{int} (\phi, \psi) = -\gamma(\phi - a)\psi^3\ , 
\end{equation}
where the dimensionless parameter $\gamma$ determines the 
strength of the interaction and where $a$ is the minimum of 
the potential $V_1$ ($\phi$).  Notice that other forms for the 
interaction potential are possible (e.g., $V_{int}\sim 
\phi^2 \psi^2$); however, the resulting behavior should be 
qualitatively the same for a fairly wide variety of choices. 

In the presence of the interaction term, the inflaton field 
$\phi$ will evolve  according to the potential 
\begin{equation}
\label{eq:vphi}
V(\phi) = {1 \over 8} \lambda (\phi^2 - a^2)^2 - \left\{ 
{\epsilon \over 2a} + \gamma \psi^3 \right\} (\phi - a) .
\end{equation}
We let the $\phi$ field be trapped in the false vacuum at the
beginning of inflation.  In the limit of nearly degenerate
vacua (small $\epsilon$) and sufficiently weak coupling
(small $\gamma$), bubbles will nucleate at a rate 
given by Eq. (\ref{eq:nucrate}), where the effective energy difference
between the vacua is given by 
\begin{equation}
\label{eq:epseff}
\epsilon_{eff} = \epsilon + 2a\gamma\psi^3\ .
\end{equation}
Notice that we have taken the small $\epsilon$ limit 
(or, equivalently, the ``thin wall limit'' see Ref. \cite{coleman})
only for the sake of obtaining analytic results; larger 
values for $\epsilon$ (and $\gamma$) will lead to similar 
behavior. 

During inflation, the equation of motion (\ref{eq:slowroll}) 
for the rolling field $\psi$ is approximately given by 
\begin{equation}
\label{eq:approx}
3H\dot\psi = - {\partial V \over \partial\psi} = - 
{\partial V_2 \over \partial\psi} + 3 \gamma (\phi - a) 
\psi^2\ , 
\end{equation}
where we have neglected the $\ddot \psi$ term in accordance
with the slow rolling approximation \cite{steinturn}. 
In the limit that $\phi$ is in the false vacuum for
essentially all of inflation (at least for purposes of 
determining the evolution of the $\psi$ field), 
we can set $\phi \simeq -\ a$ and find the equation
of motion
\begin{equation}
3H\dot\psi = - {\partial V \over \partial\psi} = - 
{\partial V_2 \over \partial\psi} - 6\gamma a \psi^2 
\equiv \fun , 
\end{equation}
where we have defined $\fun$ as the sum of the two terms above. 
The first term in $\fun$ is positive and causes the $\psi$ 
field to roll down the hill; the second term, on the other hand, 
is a negative frictional term, and later we will demand that 
this term is small enough to allow the field to roll.
The Hubble parameter is determined by
\begin{equation}
\left( {\dot R \over R} \right)^2= H^2 = 
{8 \pi G \over 3}
\left[ \rho_\phi + \rho_\psi + \rho_{rad} \right]\ , 
\end{equation}
where $\rho_\phi$ and $\rho_\psi$ are the false vacuum energy
densities of the $\phi$ and $\psi$ fields, and where $\rho_{rad}$ is
the radiation energy density.

Initially, the value of $\psi$ is small, $\epsilon_{eff}$ is small,
the nucleation rate of true vacuum bubbles $\Gamma_N (t)$ is small,
and the Universe remains in the false $\phi$ vacuum and inflates.  As
the rolling field approaches its minimum, $\psi \to \psi_f$, the value
of $\epsilon_{eff}$ becomes larger, and many bubbles (with nearly
equal sizes) of true vacuum nucleate throughout space.  Our present
Universe lies within one initially causally connected region which
experiences an inflationary epoch; the end to this inflationary period
occurs when the rolling field $\psi$ approaches the minimum of its
potential and thereby signals the old inflationary field $\phi$ to
nucleate rapidly.  Many bubbles of true vacuum nucleate simultaneously
inside the region in which the $\psi$ field is coherent (i.e., the
entire region for which we can use a single evolution equation such as
Eq.(\ref{eq:approx}) to describe the behavior of the $\psi$ field).
Our own Universe must lie within this region of coherent $\psi$.

\subsection{Constraints on the Model}

In order to obtain a successful epoch of double field inflation, we
must consider several constraints on the model parameters.  First, we
want the $\phi$ field to dominate the dynamics of the Universe and be
responsible for the inflationary epoch; hence we require $V_1(\phi) >
V_2(\psi)$.  Since $\phi \simeq -\ a$ during inflation, this
requirement becomes
\begin{equation}
\epsilon  > V_2(\psi)\ .
\end{equation}
In particular, at the beginning of inflation when $\psi$ is small (so
that $V_2(\psi)$ is near its peak), the constraint takes the form
\begin{equation}
\label{eq:inpart}
\epsilon > V_2(\psi_0)\ , 
\end{equation}
where $\psi_0$ is the value of the $\psi$ field at the beginning of the
inflationary epoch.  Given this constraint, the Hubble parameter is given by
\begin{equation}
H^2 = {8\pi G \over 3} \rho_\phi \qquad {\rm where} \qquad 
\rho_\phi\simeq \epsilon + 2\gamma a \psi^3 , 
\end{equation}
where we have made the assumption that $\phi \simeq - a$ during
inflation.

Second, in order for the coupling of the $\psi$ field to influence the
$\phi$ field and bring an end to inflation, we need the ratio $2
\gamma a \psi^3 / \epsilon$ to be sufficiently large at the end of
inflation, i.e.,
\begin{equation}
\label{eq:eta}
{2 \gamma a \psi_f^3 \over \epsilon } 
= \eta , 
\end{equation}
where $\eta$ is a dimensionless constant; in practice we require
$\eta \sim 10^{-1}$ or larger.

Third, the slowly rolling field must be able to roll despite the
frictional effect provided by the interaction term.  We must have
$\dot{\psi} > 0$, i.e.,
\begin{equation}
\label{eq:dotpsi}
- {\partial V_2 \over \partial\psi} - 6\gamma a \psi^2 = 
\fun > 0\ . 
\end{equation}

Fourth, we require that there be sufficient inflation, i.e., the total
number $N_T$ of $e$-foldings must satisfy $N_T$ $\geq$ $N_e$, where
$N_e$ is the number of $e$-foldings required to solve the original
cosmological problems ($N_e$ $\simeq$ 70). For the slow-rolling $\psi$
field, we can write the number $N_T$ of $e$-foldings in terms of an
integral and the constraint of obtaining sufficient inflation takes
the form
\begin{equation}
N_T(\psi_0 \rightarrow \psi_f)= 3 H^2
\int_{\psi_0}^{\psi_f} {d \psi \over \fun}  
\ge N_e . 
\end{equation}

Fifth, the quantum fluctuations in the field $\psi$ render its value
at any given time uncertain by the amount
\begin{equation}
\Delta \psi  > {H \over 2\pi} \ , 
\end{equation}
which leads to a constraint on the initial value $\psi_0$, i.e.,
\begin{equation}
\label{eq:quantfluc}
\Delta \psi_0  \geq {H \over 2\pi} \ , 
\end{equation}
which means that we cannot specify $\psi_0$ to an arbitrarily
precise value.

The sixth constraint is that the density fluctuations in the slowly
rolling field (see Eq.(\ref{eq:densfluc}) are not in conflict with the
observed anisotropy of the microwave background.  This requirement can
be written as
\begin{equation}
\label{eq:densfluc2}
3 H^3 / \fun \le \delta , , 
\end{equation}
where $\delta$ $\equiv$ $\delta \rho/\rho$ $\le$ $\cal O$ $(10^{-5})$
is the constraint on density perturbations \cite{cmb}.
We will also require all energy scales in the theory (e.g., the
vacuum expectation values of the $\phi$ and $\psi$ fields) to be
below the Planck scale.

\subsection{A simple example: The ramp potential}

In this subsection, we will illustrate the model of
double-field inflation by considering the simplest possible case for the
optential of the rolling field $\psi$, i.e., we will take
$\fun$ = const $>0$.  Many of the features of this simple
case aply to any version of double-field inflation. We
have chosen to present results for this simple cae
as it reveals many aspects of the double-field model
with a minimal amount of algebra.  In this 
model, the $\psi$ field will move through a potential $V_{\rm eff}(\psi)$
of the form
\begin{equation}
V_{rm eff}(\psi) = V_2(\psi) + V_{\rm int}(\psi,\phi) = V_{\rm eff}(\psi_0) 
- \fun(\psi - \psi_0) .
\end{equation}
Notice that for $\psi$ near $\psi_0$ (i.e., near the beginning of
inflation), the interaction term $V_{\rm int}$ is small and
$V_2(\psi) \sim V_{\rm eff}(\psi)$, which has a simple linear form.
With this choice of potential, the constraint that the
field $\psi$ must be able to roll (see Eq.(\ref{eq:dotpsi}) is automatically
satisfied.  Given this potential, the rolling field will begin at some initial value $\psi_0$ and roll to a final value $\psi_f$ at the end
of the inflationary epoch.  The two most restrictive constraints 
are the density perturbation constraint
\begin{equation}
\label{eq:rest1}
{3 H^3 \over \fun} \leq \delta ,
\end{equation}
and the constraint that sufficient inflation occurs
\begin{equation}
\label{eq:rest2}
{3H^2 \psi_f \over \fun} \geq N_e ,
\end{equation}
where we have taken $\psi_f >> \psi_0$.  For this theory, the fine-tuning
parameter [as defined by Eq.(\ref{eq:finpar})
can easily be evaluated and is given by
\begin{equation}
\lambda_2 = {\fun(\psi_f-\psi_0) \over (\psi_f - \psi_0)^4 } \simeq
{\fun \over \psi_f^3} ,
\end{equation}
where the subscript denotes the second field $\psi$.  Combining
the constraints of Eqs.(\ref{eq:rest1}) and (\ref{eq:rest2}), we
obtain an upper limit on the fine-tuning parameter:
\begin{equation}
\lambda_2 \leq 3 \delta^2 / N_e^3 \simeq 10^{-15} ,
\end{equation}
where we have used $\delta = 10^{-5}$ and $N_e = 70$ to obtain
the numerical value.  Thus, we obtain a fine-tuning requirement
similar to that of the standard new inflationary picture.

The fine-tuning of the potential arises in order to avoid
overproduction of density fluctuations, which are produced by the
rolling $\psi$ field (this statement is generally true for models of
inflation which involve slowly rolling fields.)\cite{afg} One might
hope that the subsequent collisions of old inflation bubbles after the
ned of the inflationary period would dominate the resultant
perturbation spectrum, especially since more energy density is
associated with the $\phi$ field than with the $\psi$ field.
Unfortunately, these bubbles are tiny compared to scales of
astrophysical interest (e.g., the scale of galaxies), which have gone
outside the horizon well before the end of inflation and have
$\psi$-field perturbations imprinted on them.  In other words, the old
inflation bubbles cannot affect struture on scales larger than the
horizon size at the end of inflation, and this size scale is much
smaller than galactic scales. Although dramatic bubble collisions can
restructure the predicted anisotropy on small scales, these collisions
cannot wipe out the unwanted large-scale perturbations produced by the
rolling field.

We can now examine the remaining constraints by writing them in terms
of the parameter $\lambda_2$ (which is constrained to be small); we
will also define a nondimensional parameter for the vacuum energy
density of the $\phi$ field, i.e.,
\begin{equation}
\label{eq:tildeeps}
\tilde\epsilon \equiv {\epsilon \over a^4} .
\end{equation}
The constraint that the $\phi$ field dominates the energy density 
of the Universe [see Eq.(\ref{eq:inpart})] can be written
\begin{equation}
\epsilon \geq V_2(\psi_0) 
\simeq V_{\rm eff}(\psi_0) \simeq \lambda_2 \psi_f^4 ,
\end{equation}
which now takes the form
\begin{equation}
\tilde\epsilon \geq \lambda_2 (\psi_f/a)^4 .
\end{equation}
The constraint that the coupling between fields is large enough to 
produce a time-dependent nucleation rate [see Eq.(\ref{eq:eta})] takes the form
\begin{equation}
\label{eq:large}
\tilde\epsilon \sim 2 \gamma (\psi_f/a)^3 .
\end{equation}
Notice that if $\psi_f \sim a$ (i.e., the vacuum expectatation values
of the two fields are comparable), then $\tilde\epsilon \sim \gamma$.

The constraint of sufficient inflation [Eq.(\ref{eq:rest2})] can now be written
\begin{equation}
\tilde\epsilon \simeq \lambda_2 {N_T \over 8 \pi} {m_{\rm Pl}^2
\psi_f^2 \over a^4} ,
\end{equation}
where $m_{\rm Pl}$ is the Planck mass and $N_T$ is the
total number of $e$-foldings ($N_T \geq N_e \simeq 70).$
If we combine this latter constraint with Eq.(\ref{eq:large}), we 
obtain the relation 
\begin{equation}
\gamma \sim {n_T \over 16 \pi} {m_{\rm Pl}^2 \over a \psi_f} \lambda_2 .
\end{equation}
Since $N_T/16\pi$ is typically of order unity, the coupling constant
$\gamma$ is larger than the (small) parameter $\lambda_2$ by the
factor $m_{\rm Pl}^2 / a \psi_f$.  In order to obtain $\tilde\epsilon
\sim \gamma \sim 1$, we must have $\psi_f \sim a$ and $a/m_{\rm Pl}
  \sim 10^{-7}$ ($ a \sim 10^{12}$GeV).  Thus, this model of
  double-field inflation can produce a reasonable scenario, provided
  that the small parameter $\lambda_2$ can be realized. Since the
  presence of such a small fine-tuning parameter is generic to
  theories of inflation which utilize slowly rolling fields (see
  Ref. \cite{afg}) for a more complete discussion), this new model is
  comparable (in temrs of fine-tuning) to existing models.

Notice that this model is described by seven parameters:  the vacuum
expectation value $a$ of the potential of the inflaton field,
the heights of the potentials $\lambda_1$ and $\lambda_2$,
the energy difference $\epsilon$< the coupling strength $\gamma$,
and finally the initial and final values $\psi_0$ and $\psi_f$ of
the rolling field.  Ideally one would like to explore fully the available range
of parameter space; such a presentation with seven parameters subject to
six constraints is byond the scope of the present paper.
However, some volume in this parameter space is allowed and will
lead to successful inflation.

Many of the features of the simple $\fun $ = const model described
above will hold in general for any version of double-field inflation. (1)
Double-field inflation will involve the seven parameters described above (in general, the final
value $\psi_f$ of the rolling field corresponds to the vacuum
expectation value of the $\psi$ field).  (2) Large-scale perturbations
(i.e., on the scale of the present horizon down to the scale of galaxies)
will be produced in a manner analogous to that of new inflation.
These perturbations will not be erased through the nucleation and
subsequent thermalization of bubbles of the $\phi$ field
(these bubbles have size scales comparable to the
horizon at the end of inflation, i.e., much smaller than the scale
of galaxeis).  (3)  To avoid overproduction of density perturbations
on large scales, the potential of the slowly rolling field $\psi$
must be very flat, with a fine-tuning parameter $\lambda \sim 10^{-15}$.

We have also considered more realistic choices for the potential
of the $\psi$ field in double-field inflation.  For example,
we have examined a potential $V_2(\psi)$ of the Coleman-Weinberg
form \cite{colewein}, i.e.,
\begin{equation}
V_2(\psi) = {1 \over 2} B \sigma^4 + B \psi^4 [{\rm ln}(\psi/\sigma)^2 - 
{1 \over 2} ],
\end{equation}
where $\sigma$ is the vacuum expectation value of the $\psi$ field
and $B$ characterizes the flatness of the potential and is
analogous to the parameter $\lambda_2$ defined above.  A discussion of
double-field inflation with a Coleman-Weinberg potential is given in
the Appendix.  We find that successful double-field inflation can
occur with this potential, although the constraints on the model are
even more restrictive than in the simple case outlined above.  In
particular, the constraint that the $\psi$ field can roll initially
[see Eq.{\ref{eq:dotpsi})] implies that the coupling parameter $\gamma$ must
  be much smaller than unity; the constraint of Eq.(\ref{eq:large})
then implies that $\tilde\epsilon$ must also be small for this case.
For example, if we take $\sigma \sim a \sim m_{\rm Pl}$, we 
find that $B \leq 3 \delta^2/8 N_T^3 \simeq 10^{-15}$
and that $\tilde\epsilon \sim \gamma \sim B$.  Thus,
fine-tuning arises in this model.  Alternatively, a model with
two different inherent mass scales (similar to the case of
schizons \cite{hillross} or axions \cite{weinwilc}) may
provide the necessary flat potential \cite{nat}.

\subsection{Evolution of the Probability Function}

Once the necessary constraints are satisfied, we can 
solve for the evolution of the Universe.  In particular, 
we can find the probability of finding the inflaton 
field $\phi$ in its false vacuum state. 
This probability can be written [see Eq.(\ref{eq:prob}) and
Ref. \cite{guthwein}]
\begin{equation}
p(t) = \exp \left\{ - {4 \pi \over 3}\ {A \over \chi_0^4}\ 
\int_{t_i}^t \chi_0 d t^\prime e^{-{S_E}({t^\prime})}
[ 1 + \eta \bar\psi^3 ]^{-3/2} \right\} , 
\end{equation}
where $S_E$ is the Euclidean action and is given by 
\begin{equation}
\label{eq:action}
S_E = S_0 \, [ 1 + \eta \bar\psi^3 ]^{-3} , 
\end{equation}
where we have defined $\bar\psi \equiv \psi/\psi_f$ and
where the dimensionless constant $\eta$ is given by 
Eq.(\ref{eq:eta})
[notice that in this present notation, $\epsilon_{eff} = \epsilon 
( 1 + \eta \bar\psi^3 )$].

Since the number of $e$-foldings (of the scale factor) is the
relevant time variable for inflation, we change variables according to
\begin{equation}
d\tau = \chi dt = \chi_0 (1+\eta \bar\psi^3)^{1/2}dt ;
\end{equation}
we can then write the probability as
\begin{equation}
\label{eq:probnew}
p(\tau) = \exp \left[ -\alpha \int_{0}^{\tau} d \tau 
e^{-S_0 [ 1 + \eta \bar\psi^3 ]^{-3}} ( 1 + \eta \bar\psi^3 )^{-2} 
\right] ,
\end{equation}
where we have defined
\begin{equation}
\label{eq:alpha}
\alpha \equiv {4 \pi \over 3} {A \over \chi_0^4} ,
\end{equation}
which will generally be of order 1 (see Ref.\cite{callan}). 
Equation (\ref{eq:probnew}) 
can be written as a differential equation, 
\begin{equation}
\label{eq:diff}
{dp \over d\tau} = - \, p(\tau) \, \alpha \, 
e^{-S_0 [ 1 + \eta \bar\psi^3 ]^{-3}} 
[ 1 + \eta \bar\psi^3 ]^{-2} , 
\end{equation}
which can be integrated numerically once we have solved the evolution
equation (\ref{eq:approx}) for the $\bar\psi$ field. As an example, we
will consider the $\fun = $const model presented in the preceding
subsection; for this case, the equation of motion of the $\bar\psi$
field takes the form
\begin{equation}
\label{eq:b}
(1+\eta \bar\psi^3) {d\bar\psi\over d\tau} = 
{\fun \over 3 \chi_0^2 \psi_f} = {\lambda_2
\over 3 \chi_0^2} \equiv {\cal B} ,
\end{equation}
where we have defined a new constant ${\cal B}$ [in the form of
Eq.(\ref{eq:b}), the equation of motion can be easily integrated].

\begin{figure}
\centerline{\includegraphics[width=5.5in]{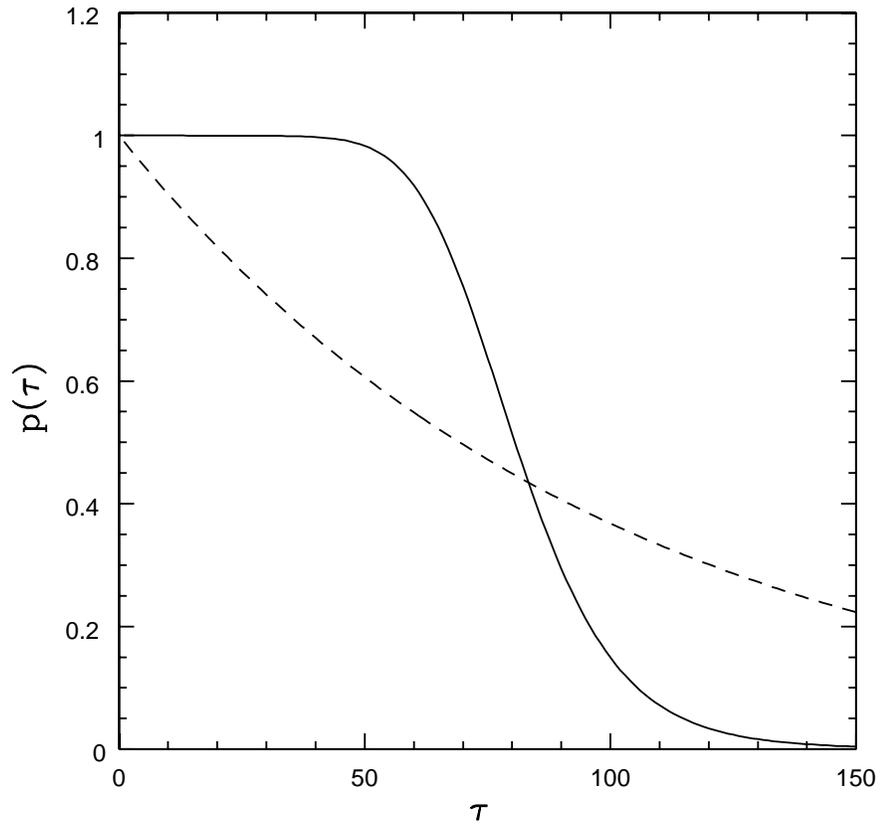}}
\caption{Probability $p$ of a point in space being in the false
vacuum as a function of nondimensional time $\tau$.  Solid curve
shows the double-field inflation model of Sec. III; for comparison,
the dashed curve shows the case of constant nucleation efficiency
(as in old inflation).}
\end{figure}

With this formulation of the problem, we must specify three parameters
($S_0, \eta,$ and ${\cal B}$) to determine the evolution of the probability
function (we have taken $\alpha=1$). The initial value $S_0$ of the
action must be large enough to make the initial nucleation rate small,
but small enough to allow for a sufficiently rapid nucleation rate at
the end of inflation [see Eq.(\ref{eq:action})]; we thus require
$S_0 \sim 10$. The interaction strength is given by $\eta$, which must
be large enough to affect the evolution of the Universe but small
enough not to dominate the dynamics; we thus require $\eta \sim
0.1-1$.  The constant ${\cal B}$ essentially determines the number of
$e$-foldings [see Eq.(\ref{eq:b})], so we must have ${\cal B} = O(1/N_e)$. If
we choose $S_0 =12$, $\eta=0.4$, and ${\cal B}=0.02$, the resulting
probability evolution function $p(\tau)$ is shown in Fig. 2.  By
choosing parameters appropriately, we can arrange to have nucleation
efficiency $\beta$ [see Eq.(\ref{eq:beta})] subcritical initially and
thereby obtain sufficient inflation.  Since $\beta$ is now time
dependent, we can also have $\beta$ supercritical for the latter part
of the inflationary epoch and thereby allow for percolation of the
true-vacuum bubbles.  Notice also that the probability function
$p(\tau)$ is much more like a step function (see Fig. 2) than for the
case of old inflation (i.e., constant nucleation efficiency). This
result implies that most of the bubbles of true vacuum which are
nucleated will have sizes comparable to the horizon scale at the end
of the inflationary epoch; since this size scale is small compared to
size scales of astrophysical interest (e.g., the scale of galaxies),
the only relevant density fluctuations produced by this inflation will
result from the rolling field $\psi$ and not from the inflaton field
$\phi$ if the phase transition is infinitely sharp. Notice, however,
that the end of the phase transition is not infinitely sharp; the
width of the phase transition shown in Fig. 2 is approximately 20
$e$-foldings.  We have not calculated the details of the end of this
phase transition; we leave this study of the thermalization for future
work. Notice, however, that the nucleation of the inflaton field may
generate additional large-scale structure which may explain some of
the features we observe today.

\section{Discussion}

We have studied inflationary scenarios which (ab)use two 
coupled real scalar fields; the coupling between fields
can lead to a time dependent nucleation rate.  We thus
obtain a successful inflationary scenario which ends 
through a first order phase transition, i.e., through
the nucleation of true vacuum bubbles in the sea of
false vacuum.  The required degree of flatness in the 
potential of the rolling field $\psi$ is comparable to
that required in `new inflation', i.e., 
$\lambda$ = $\cal O$ $(10^{-15}$).  This present model
is thus comparable in success to existing models, but 
occurs in a different manner and may contain some 
advantages. For example, the inflationary epoch ends through 
the process of nucleation and topological defects such as 
cosmic strings \cite{kibblevil} can form at the end of the phase 
transition (provided that the potential of 
the $\phi$ field is complex).  

``Extended'' inflation \cite{lastein} also revives some
of the aspects of the ``old'' inflation models in that
the inflation takes place at a supercooled first-order phase transition.
The essential difference from old inflation is that
gravity is described not by general relativity, but by
Brans-Dicke \cite{bransdicke} theory.  Extended
inflation also provides a time-dependent nucleation efficiency;
however, the time dependence is achieved through a time-dependent
Hubble parameter [see the denominator of Eq.(\ref{eq:beta})]
rather than through a time-dependent nucleation rate [the
numerator of Eq.(\ref{eq:beta})]. Studies of bubble
nucleation, collisions, and percolation \cite{weinberg,lsb}
restrict the allowed parameters of the model and the potential
of the coupled field also must be fine-tuned (see Ref. \cite{afg}).
A generalized version of extended inflation (``hyperextended
inflation'' \cite{steinacc}) utilizes more complicated couplings
of the rolling field to gravity to obtain a time-dependent
Hubble parameter (and hence a time-dependent $\beta$).

The specific model of double field inflation 
presented in this paper can produce a ``successful'' 
inflationary epoch.  However, the theory must contain 
a very small parameter (namely $\lambda_2$ $\sim$ 10$^{-15}$)
in order to satisfy constraints on density perturbations.
Although this particular (highly simplified) model is unlikely 
to provide the ultimate inflationary scenario, the concept of 
a time-dependent nucleation rate provides a very promising 
mechanism.  Freese, Frieman, and Olinto \cite{nat} proposed
a model using pseudo Nambu-Goldstone bosons that naturally
involves two disparate mass scales and thus gives 
very flat potentials without any fine-tuning of parameters;
potentials such as these are ideal for the rolling field
in double-field inflation.

{\it Note added in proof.} After the completion of this
paper, we discovered that A. Linde has simultaneously
suggested the possibility of a time-dependent nucleation
rate through the coupling of scalar fields [Report No.
CERN-TH.5806/90,1990(unpublished)].

\section*{Acknowledgements}
 
We would like to thank Alan Guth for many useful discussions and for
his wise counsel; we also thank Premi Chandra for stimulating
conversations about time-dependent nucleation rates arising in
condensed matter physics due to long-range interactions.  K. F.
acknowledges support from the Alfred P. Sloan Foundation (Grant No.
26722), the National Science Foundation through the Presidential Young
Investigator Program, and NASA (Grant. No. NAGW-1320).

\section*{Appendix:  The Coleman-Weinberg Case} 

In this appendix, we will consider a more realistic model
of double-field inflation using a Coleman-Weinberg \cite{colewein}
form for the
potential of the slowly rolling field, i.e., we will take
\begin{equation}
V_2(\psi) = {B \sigma^4 \over 2} + B\psi^4 \left[ \ln \left({\psi^2
\over \sigma^2} \right) - {1 \over 2} \right]\ .
\end{equation}
The potential for the $\phi$ field is still described by
Eq.(\ref{eq:phipot}) with the interaction term of
Eq.(\ref{eq:intpot}).  The $\psi$ field starts rolling at some initial
value $\psi_0 \geq H/2 \pi$ and finally reaches its stable minimum
at $\psi_f = \sigma$.  With this choice of potential,
the equation of motion (\ref{eq:approx}) becomes
\begin{equation}
3H\dot\psi = -4 B \psi^3 \ln \left( {\psi^2 \over \sigma^2} \right)
- 6 \gamma a \psi^2\ .
\end{equation}

We will consider the constraints for this potential.
We will consider the special case where $\sigma=a$ and
will define $\tilde\epsilon = \epsilon/a^4$ as in Eq.(\ref{eq:tildeeps}).
With these restrictions, the requirement that the $\phi$ field dominate
the dynamics of the Universe and cause an inflationary
epoch becomes
\begin{equation}
\label{eq:tilde}
\tilde\epsilon \geq B/2 .
\end{equation}
In order for the coupling to the $\psi$ field to influence 
the $\phi$ field  we require that
\begin{equation}
\gamma \geq \tilde\epsilon /20 ,
\end{equation}
where we have taken $\eta = 0.1$ [see Eq. (\ref{eq:eta})].  The third
constraint Eq. (\ref{eq:dotpsi}), the requirement that $\dot\psi > 0$
in order for the $\psi$ field to enter a slow rolling epoch, becomes
\begin{equation}
\lambda_B(\psi_0/a) \geq 6 \gamma ,
\end{equation}
where $\psi_0$ is the initial value of the $\psi$ field
and where we have defined $\lambda_B \equiv 4B {\rm ln}(a^2/\psi_0^2)$.
The condition of sufficient inflation then becomes
\begin{equation}
\label{eq:apsuff}
N (\psi_0 \to \psi_f) \simeq {3 H^2 \over 2 \lambda_B}
\left[ {1 \over \psi_0^2} - {1 \over \psi_f^2} \right] \sim
{3 H^2 \over 2 \lambda_B} {1 \over \psi_0^2} \geq N_e , 
\end{equation}
where we have assumed that the final value $\psi_f >> \psi_0$ and
where $N_e$ is the required number of $e$-foldings.  The quantum
fluctuation constraint [Eq. (\ref{eq:quantfluc})]) remains the same.
Finally, the constraint that the density fluctuations are sufficiently
small [see Eq.(\ref{eq:densfluc2})] takes the form
\begin{equation}
\label{eq:apdens}
{3 H^3 \over \lambda_B \psi_0^3} \leq \delta .
\end{equation}
The coupled constraints of Eqs. (\ref{eq:apsuff}) and
(\ref{eq:apdens}) can be combined to obtain a bound on the parameter
$\lambda_B$:
\begin{equation}
\label{eq:lamb}
\lambda_B \leq {3 \delta^2 \over 8 N_e^3} .
\end{equation}
The numerical value of the right-hand side of Eq.(\ref{eq:lamb}) is of
order $10^{-14}$; we thus obtain a ``fine-tuning'' requirement which
is comparable in magnitude to that of new inflation.  Let us now
saturate the constraint of Eq. (\ref{eq:tilde}); i.e., we will take
$\tilde\epsilon = B/2$.  If we then consider the specific
case $a=m_{pl}$ and define $x=\psi_0/a=\psi_0/m_{pl}$
(where the dimensionless parameter $x$ must be less than unity),
we can write the remaining constraints in the form
\begin{equation}
\label{eq:apfirst}
\gamma \geq B/40,
\end{equation}
\begin{equation}
\label{eq:apsecond}
4Bx {\rm ln}(1/x) \geq 3 \gamma ,
\end{equation}
\begin{equation}
\label{eq:apthird}
{\pi \over 4 N_e} \geq x^2 {\rm ln}(1/x) .
\end{equation}
If we saturate the third constraint [Eq.(\ref{eq:apthird})] and solve
for $x$ we obtain $x=0.064$ (where we have taken $N_e=70$).  With this
value for $x$, the remaining two constraints [Eqs.(\ref{eq:apfirst})
and (\ref{eq:apsecond})] confine the ratio $B/\gamma$ to the range
\begin{equation}
\label{eq:number}
4.3 \leq B/\gamma \leq 40 .
\end{equation}
With these values of the parameters, the quantum fluctuations
constraint [Eq.(\ref{eq:quantfluc})] is satisfied.  Thus, there exists
a region of parameter space which allows successful inflation with two
coupled scalar fields.  Notice, however, that Eq.(\ref{eq:lamb})
constrains $\lambda_B$ (hence $B$) to be very small.  In our specific
example, $\tilde\epsilon = B/2$ (by assumption) and $\gamma$ is within
an order of magnitude of $B$ [by Eq.(\ref{eq:number})], so that both
$\tilde\epsilon$ and $\gamma$ are also very small in this case.

\end{document}